# Tunable plasmonic devices by integrating graphene with ferroelectric nanocavity

Junxiong Guo[1]*, Shangdong Li[2,3]*, Jianbo Chen[1,4], Ji Cai[1], Xin Gou[1], Shicai Wang[2], Jinghua Ye[1], Yu Liu[5], Lin Lin[1,2]*

[1]School of Electronic Information and Electrical Engineering, Chengdu University, Chengdu 610106, China

[2]School of Electronic Science and Engineering (National Exemplary School of Microelectronics), University of Electronic Science and Technology of China, Chengdu 610054, China

[3]School of Electronics and Information Technology (School of Microelectronics), Sun Yat-sen University, Guangzhou 510006, China

[4]College of Material Science and Engineering, Sichuan University, Chengdu 610064, China

[5]School of Integrated Circuits, Tsinghua University, Beijing 100084, China

*Corresponding authors.
E-mail: guojunxiong@cdu.edu.cn (J.X. Guo); lishd8@mail2.sysu.edu.cn (S.D. Li) and linlin@std.uestc.edu.cn (L. Lin)

**Abstract**

Graphene plasmons are able to become the fundermental of novel conceptual photonic devices, resulting from their unique characteristics containing excitation at room temperature and tunable spectral selectivity in different frequencies. The pursuit of efficiently exciting and manipulating graphene plasmons is necessary and significant for high-performance devices. Here, we investigate graphene plasmon wave propagating in ferroelectric nanocavity array. We experimentally show that the the periodic ferroelectric polarizations could be used for doping graphene into desired spatial carrier density patterns. Based on a theoretical model that considers periodic ununiform conductivity across graphene sheet, the simulation results show surface plasmon polaritons (SPP) in graphene can be excited by an incident light in a similar way to the excitation of photonic crystal resonant modes. The graphene SPP resonance can be tuned from ~720 to ~1 000 $cm^{-1}$ by rescaling the ferroelectric nanocavity array, and from ~540 to ~780 $cm^{-1}$ by dynamically changing the applied gate voltage. Our strategy of graphene carrier engineering to excite SPP offers a promising way for large-scale, non-destructive fabrication of novel graphene photonic devices.

**Keywords:** graphene, tunable device, ferroelectric, nanocavity, surface plasmon polariton

## 1. Introduction

A key feature of graphene is that its plasmons can be externally manipulated and used to resonantly enhance light-matter interactions.[1-3] The diversity of applications for electro-optical modulation,[4, 5] bio-sensing,[6, 7] micro-spectrometry,[8, 9], tunable photodetection,[10-13] have been realized using graphene plasmon polaritons due to the low optical losses and high confinement of electromagnetic field,[14-16] compared with conventional plasmonic materials. The exploits of graphene intrinsic plasmons, generating from collective charge oscillations of two-dimensional electron liquid, have been demonstrated as tunable infrared and terahertz detectors using patterned graphene, such as graphene antidot lattices,[17-20] graphene nanoribbons (GNR) with different stripe widths,[10, 21] graphene nanodisks,[13] and hybrid dimensions.[22] More importantly, both the wavelength and the amplitude of the plasmons in GNR can be dynamically controlled by an external voltage, which were detected by infrared nano-imaging experiments.[23, 24] However, conventional pattern methods such as reactive ion etching (RIE) would the inevitably introduce edge-disorder in graphene, which would increase carriers scattering and reduce the quality-factor of the corresponding plasmonic devices.[25-28] Development of controllable and stable doping of graphene with undamaging techniques for plasmons manipulation is therefore highly desirable.

Ferroelectric thin films, with reversibly printable nanoscale domains (generally 0.5-10 nm wide),[29-31] provide an ideal platform for precisely doping graphene into desired spatial patterns to study surface plasmon polaritons (SPP) wave manipulation. In this article, we use a uniformly down-polarized $BiFeO_3$ (BFO) thin film with periodic nanoscale cavity to excite and confine graphene plasmons. Dynamical Raman signals are used to monitor graphene carriers' behavior on patterned BFO substrates under different gate voltages. The graphene carrier density is estimated to reach $2.2 \times 10^{13}$ cm$^{-2}$ at bias-voltage of −4 V. Numerical simulations are calculated using finite element method, indicating that the resonance effect of exciting SPP waves in graphene to incident lights occurs at edges of ferroelectric nanocavity. With scaling the period of patterned ferroelectric nanocavity, the position of resonance peaks can be readily modulated from ~720 to ~1 000 cm$^{-1}$. Our devices also feature a tunable resonance in transmission spectra by directly varying the applied gate voltage.

## 2. Results and Discussion

### 2.1. Concept and fabrication of plasmonic device

Figure 1a-e shows the concept and fabrication of our tunable plasmonic device based on stacking of graphene with nanopatterned BFO thin film. The fabrication processes are detailed discussed in Methods section. The BFO thin films, epitaxially grown on $SrTiO_3$ (STO)

substrates with a conductive (La,Sr)MnO₃ (LSMO) bottom layer, were etched to nanocavity arrays with periodicity ($a$) of 100 to 220 nm. Then, the patterned BFO thin films were switched using a large-scale, green water-printing method.[32] After that, the CVD-growth graphene sheets were transferred onto the prepared BFO with uniformly downward polarization (UDP-BFO) thin films with nanocavity array. Figure 1f is the scanning electron microscopy (SEM) image of the integration of graphene with BFO circular nanocavity substrate arranged in a rectangle lattice. Here, the diameters of nanocavity ($d$) and periodicity of a unit are 120 and 180 nm, respectively.

The piezoresponse force microscopy (PFM, Figure 2a) image shows the phase mapping, indicating that the out-of-plane component of BFO polarization is uniformly downward. The transmission electron microscopy (TEM, Figure 2b) image presents the switching behaviors over the whole BFO film within the volume. The enlarged view (Figure 2c) shows the Fe atoms (fake yellow spheres) move upward, indicating that the downward polarization is distributed across the whole BFO thin film. The Raman spectrum of the transferred CVD graphene (Figure 2d) shows a small intensity ratio of D-band to G-band (< 2%), indicating the high-quality graphene on patterned BFO.[33, 34]

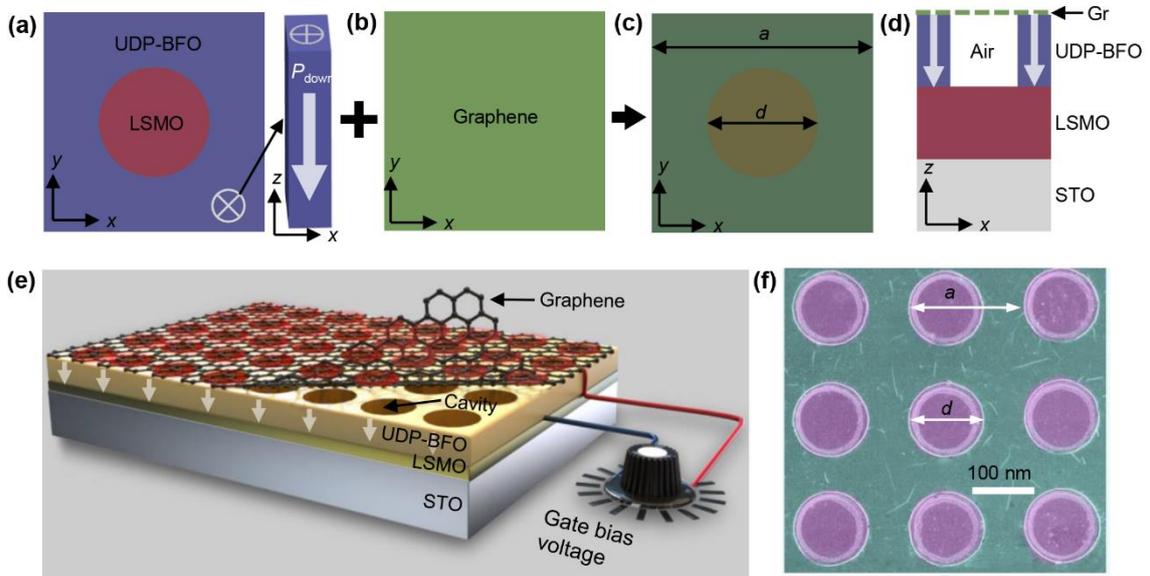

**Figure 1. Conceptual design and fabrication of plasmonic device.**
(**a-c**) Experimental scheme. Epitaxial BFO thin film was etched to nanocavity array and uniformly switched to downward polarization ($P_{down}$) in panel (**a**). CVD growth graphene in panel (**b**) was transferred onto the uniformly down-polarized BFO (UDP-BFO) to assemble the device in panel (**c**). (**d**) Schematic cross-section in a unit of the device. $a$ and $d$ represent the period of the unit and the radius of nanocavity. (**e**) Architecture of designed plasmonic device using graphene/ferroelectric cavity array. The bottom electrode layer of LSMO is used to both switch the ferroelectric domain and apply gate bias voltage. (**f**) SEM image (top view, false color) of the fabricated device consisting of graphene and ferroelectric nanocavity array ($d$ = 120 nm and $a$ = 180 nm). The arrows of schematics in panels (**a, d, e**) represent the out-of-plane component of the polarization.

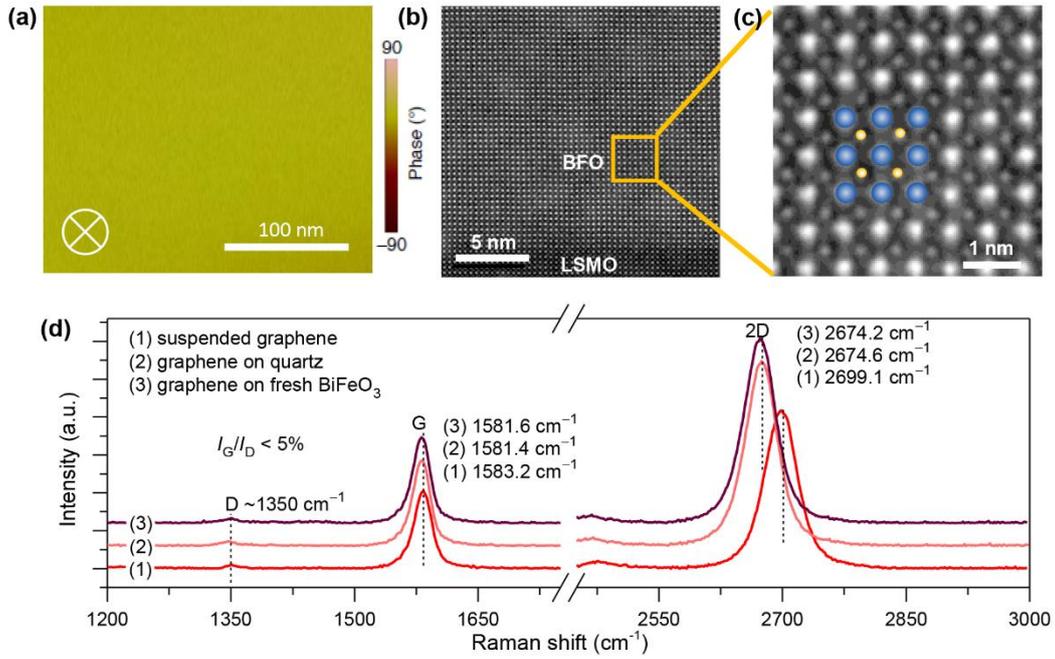

**Figure 2. Structure charaterization of device.**
(**a**) PFM images for BFO thin film after exposuring within acidic solution (pH = 3). (**b**) TEM image for epitaxial BFO thin film on LSMO. (**c**) The enlarged view of BFO film in panel (**b**). (**d**) Raman shifts for suspended graphene, and graphene on fresh BFO and quartz substrates.

**2.2. Tuning of graphene carrier behavior**

Unlike localized plasmonic resonances supported by confined structures such as graphene disks,[13] the graphene in our designed device is continuous. Therefore, the excited graphene plasmonic modes are actually propagating SPP according to the previous theoretical works.[5, 20] We expect that the periodic nanoholes in the BFO film modulate the carrier density distribution in the continuous graphene, contributing to the period Bragg scattering along the propagation of surface plasmon waves in the graphene. As a result, various SPP modes can be excited by an incident light in a way similar to the excitation of photonic crystal resonant modes. Compared the graphene carrier modulation by etched methods, such as nanodots,[17-19] our designed devices have a feature of that the graphene is intact and without secondary processing, which avoids the likely carrier mobility degradation due to the edge roughness of the etched graphene.

We first examined Raman shifts in graphene because the shifts of Raman G-band frequency could act as a spatially resolved probe monitoring graphene carrier-density.[33] The characterization processes are detailed shown in Methods section. Figure 3a schematically shows the experimental setup for the non-contact detection of graphene carrier behaviors by combining atomic force microscope (AFM) tip with Raman technique. As shown in Figure 3b, the map presents that the lower and higher Raman G-band frequencies of suspended graphene on nanocavity and graphene on UDP-BFO under zero-bias gate voltage, respectively. The periodicity of Raman G-band frequency shifts from lower to higher values implying that the graphene features a sharp transition from lower to higher carrier concentration corresponding to etched hole and UDP-BFO locations.

We further extract the average Raman G-band frequency of graphene on UDP-BFO and suspended graphene on nanocavity (Figure 3c). The peak positions of G-band (POG, $\omega_G$) for graphene on UDP-BFO and nanocavity are 1594.5 cm$^{-1}$ and 1583.6 cm$^{-1}$, respectively. The significant difference of POG ($\triangle\omega$) of graphene on UPD-BFO to pristine graphene in vacuum ($\omega_G$ = 1580 cm$^{-1}$, ref. [35]) reveals the UDP-BFO doped graphene has an ultrahigh carrier density, bahaving typically *p*-doped characteristics, which has reported previously.[30, 36, 37] The small vibration of POG for graphene on ferroelectric cavity may relate to the absorbed H$^+$/OH$^-$ ions during fabrication and the intrinsic crystal defects.

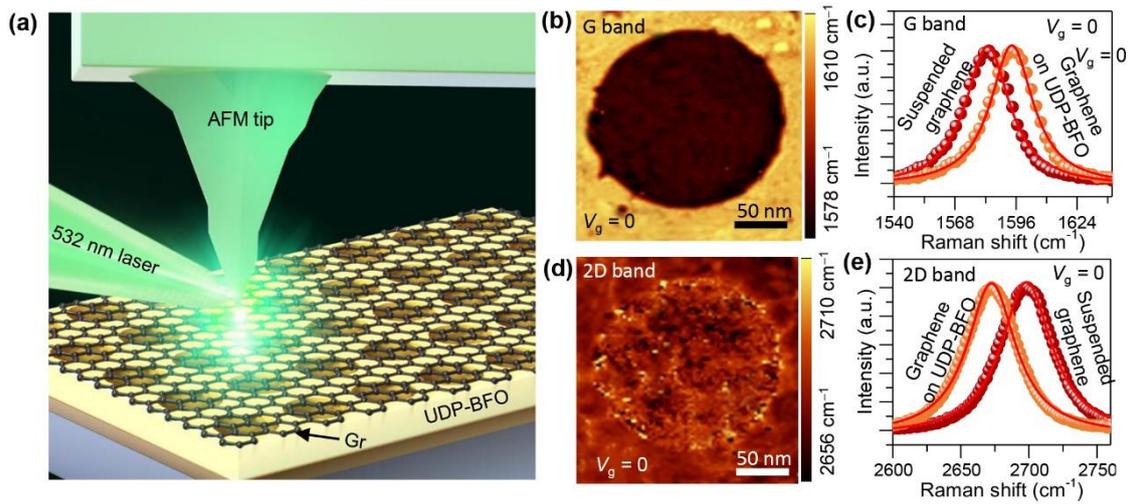

**Figure 3. Spatially controlling of graphene carrier density.**
(**a**) Experimental schematic of tip-enhanced Raman spectrometer for non-contact monitoring of graphene carrier density. (**b**) Raman mapping of graphene G-band frequency. (**c**) Averaged data plots of graphene G-peak. (**d**) Raman mapping of graphene 2D-band frequency. (**e**) Averaged data plots of graphene 2D-peak. The solid lines in panels (**c**, **e**) correspond to Lorentz fitting results to the experimental data (light orange spheres for the graphene on the UDP-BFO and dark red spheres for the suspended graphene on the nanocavity). The Raman shifts were charaterized under zero-bias gate volatge.

The Raman 2D-band frequencies were also measured, which derives from a second-order, double-resonant (DR) Raman scattering mechanism.[38] Figure 3d,e shows the measured Raman 2D-band frequency maps under zero-bias gate voltage and corresponding extracted average data, respectively. It shows the peaks for graphene on UDP-BFO and nanocavity are 2672.7 cm$^{-1}$ and 2699.1 cm$^{-1}$, respectively. This blue-shift (26.4 cm$^{-1}$) is too large to be obtained by electron/hole dopants by comparing the dependence of doping on shift in the 2D-band, which is generally ~10% to ~30% of G-band shift.[39] Herein, we attribute this abnormal blue-shift of 2D band to the strain effect induced by the BFO nanocavity in the substrates.

We now consider the evolution of POG dependence of gate voltage. The bias gate voltage can be directly applied on the device via the conductive LSMO bottom electrode (Figure 4a and Figure S1 in Supporting Information). Figure 4b plots the representative gate-voltage-dependent Raman POG of graphene on UDP-BFO thin film, supporting the above statement of graphene *p*-doped behaviors on UDP-BFO film. However, the POG of suspended graphene on cavity just has a small vibration, revealing the behavior as near-pristine graphene characterizations.

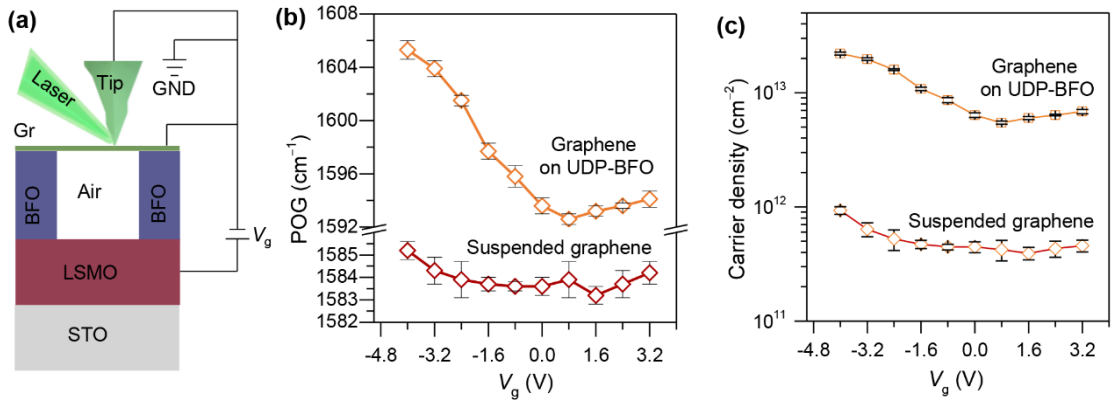

**Figure 4. Gate-dependence of graphene carrier density.**
(**a**) Experimental setup. (**b**) Position of graphene G-band as function of applied gate voltage. (**c**) Estimated graphene carrier density as function of applied gate voltage. Error bars indicate the S.D. of the frequency (**b**) and corresponding estimated graphene carrier density (**c**) within each domain.

The doping level of graphene can be evaluated through the POG ($\omega_G$) of doped graphene to pristine graphene by equation (1),[35]

$$\omega_G - 1580 \text{ cm}^{-1} = \left(42 \text{ cm}^{-1}/\text{eV}\right) \times |E_F| \quad (1)$$

where $E_F$ is the Fermi level of graphene. The carrier concentration (*n*) can be estimated from the following Fermi energy,[33]

$$E_F = \hbar v_F \sqrt{\pi n} \quad (2)$$

where $v_F$ is the Fermi velocity of $1.1 \times 10^6$ m s$^{-1}$. Figure 4c shows the graphene carrier density on UDP-BFO can be easily tuned from ~$5.5 \times 10^{12}$ cm$^{-2}$ to ~$2.2 \times 10^{13}$ cm$^{-2}$ by varying the gate voltage from +3.2 to −4.0 V. The actual carrier density of graphene can be obtained by using the graphene quantum capacitance and the geometrical gate capacitance (see Section S1 in Supporting Information). We obeserve that the estimated carrier density from Raman shifts is slightly higher than actually obtained carrier density. For example, the graphene has a carrier density of ~2.2×10$^{13}$ cm$^{−2}$ monitored by Raman shifts, while the actual carrier density of graphene is ~2.0×10$^{13}$ cm$^{−2}$ calculated by electrical characterizations (Figure S2 and Figure S3 in Supporting Information). The slight difference is reasonable because of the strain effect on Raman characterization. On the other hand, the suspended graphene on cavity maintains an order of 10$^{11}$ cm$^{−2}$, which is close to that of the graphene on SiO$_2$ substrate. It has suggested that this doping of suspended graphene depends on the charged impurities and the bended intensities.

## 2.3. Excitation and confinement of graphene plasmons

Based on above-discussed interesting carrier density pattern spatially controlled by the ferroelectric nanocavity, we can easily achieve a desired conductivity patterns because of the high-dependence of chemical potential ($\mu_c$ = $E_F$) to graphene's dynamical conductivity. According to the theoretical prediction,[5] this fabricating inhomogeneous permittivity of patterned ferroelectric spacer dielectric provides a promising technique to efficiently manipulate SPP in graphene.

Figure 5a presents the three-dimensional model of simulated electric-field ($E$-field) distribution of our designed plasmonic device. Here, the diameter of nanocavity $d$ and periodicity of unit $a$ are 90 and 180 nm, respectively. The chemical potentials of graphene on UDP-BFO and suspended graphene on nanocavity are assumed by extracting our Raman experiment data. For example, 604 meV of UDP-BFO-doped graphene and 124 meV of suspended graphene on nanocavity can be achieved according to equation (1) when the applied gate voltage is −4 V. The dynamical conductivity values calculated from Kubo formula[5] are, $\sigma_{g,1}$ = 0.0054 + $i$0.479 mS (UDP-BFO-doped graphene) and $\sigma_{g,2}$ = 0.0085 + $i$0.0801 mS (suspended graphene on nanocavity), with room temperature of 300 K and charged particle scattering rate of 5.32 meV and 1.09 meV, respectively, where the carrier mobility is obtained by the Drude model as $\mu$ = ($en\rho$)$^{−1}$ (see Figure S2). Our numerical simulations clearly reveal that the $E$-field's intensity near the nanocavity edge is evidently higher than that in the rest of regions, suggesting that the plasmon waves are localized around the edges of nanocavity, as shown in Figure 5b and corresponding enlarge view Figure 5d. The plots of $E$-field enhancement ($|E|/|E|_{max}$, Figure 5e,f) extracted from the cross-section of device (Figure 5c) further support that the plasmon waves are highly localized in graphene on patterned BFO nanocavity. When the highly confined plasmon wave vector matches to the incident wavelengths, the light-matter interaction is considerably enhanced and the resonance peaks could be easily tuned.

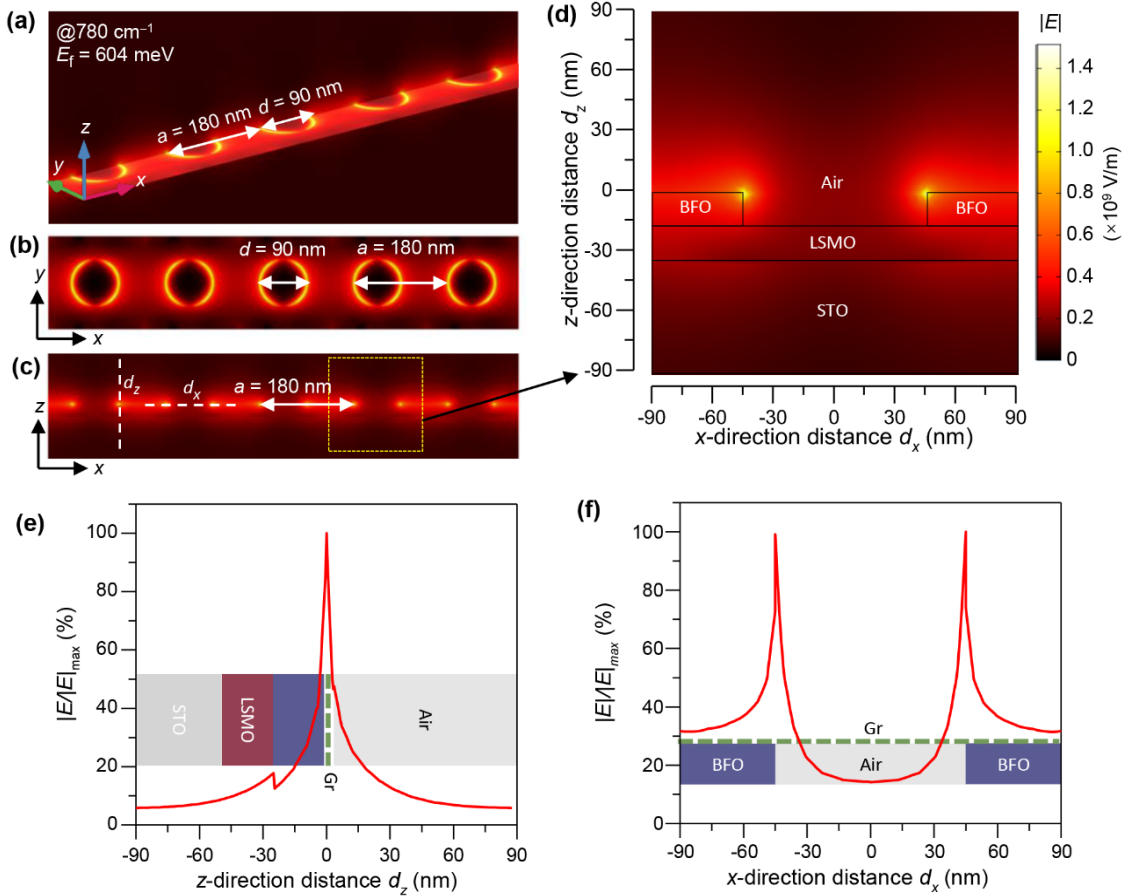

**Figure 5. Graphene plasmons excitation and confinement by ferroelectric cavity.**
Simulated results of electric-field (*E*-field) distribution under resonant frequency with overall view (**a**), top-view (**b**) and cross-section view (**c**). (**d**) Simulated *E*-field distribution corresponding to the enlarged view of panel (**c**) within yellow dashed square. Percentage of electric-field intensity confined within volume extending *z*-distance (**e**) and *x*-distance (**f**) outside the device, corresponding to the perpendicular and parallel dashed lines in panel (**c**).

For two-demensial systems, the plasmon dispersion relation can be expressed as equation (3), [15]

$$\omega_{pl} = \left( \frac{e^2 E_F q}{2\pi \hbar^2 \varepsilon_0 \varepsilon_r} \right)^{1/2} \tag{3}$$

where $\varepsilon_0$, $\varepsilon_r$ represent the vacuum permittivity and spacer dielectric constant, respectively. The resonant frequency ($\omega_{pl}$) depends on the wave vector $q$ and Fermi level of graphene $E_F$. Thus, there are at least two major routes for manipulating plasmon waves in our conceptual device: (i) passive tuning of graphene plasmons by reconstructing ferroelectric nanocavity diameter $d$ and periodicity $a$ to match the plasmon wave vector, and (ii) active tuning of graphene plasmons by dynamically changing the Fermi level or carrier concentration via electrostatic gating. [9, 13]

## 2.4. Tunable spectral response

We now turn to discuss the spectral response in the infrared region of device. The extinction spectra in transmission, $1-T/T_0$, measured on as fabricated, graphene/ferroelectric cavity with diameter $d$ from 80 to 120 nm and identical periodicity a of 180 nm are presented in Figure 6a. The inset of Figure 6b shows a schematic of the experimental characterization setup for such a device. Due to the identical periodicity of nanocavity, the resonant frequencies of as-prepared devices are around ~780 cm$^{-1}$. When the diameter of nanocavity exceeds 100 nm, the resonant transmission band splits to two peaks. To realize this frequency splitting instead of frequency resonant selectivity, we further simulated the $E$-field distribution in our device. Our simulations (Figure 6b) reveal that plasmons excited by the "ring-like" graphene conductivity patterns. Consequently, the interactions between two neighboring nanocavity can not be ignored when the diameter of nanocavity is too large (bottom panels of Figure 6b).

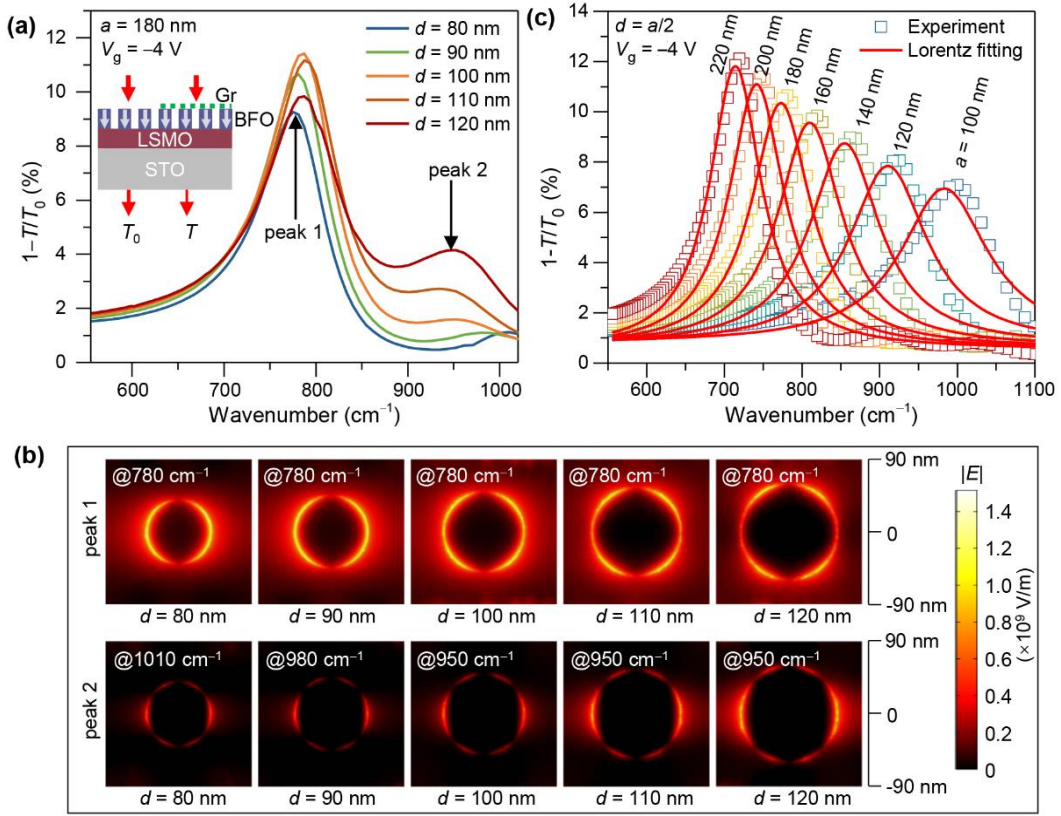

**Figure 6. Geometric-tunable infrared filters.**
(**a**) Extinction in transmission, $1-T/T_0$, in device of graphene/ferroelectric nanocavity array with identical period ($a$ = 180 nm) but different diameters ($d$ ranges from 80 to 120 nm). (**b**) Simulated electric-field distribution of graphene under first and secondary resonant frequencies, corresponding to panel (**a**). (**c**) Tuning of infrared extinction spectra by varying the period of the nanocavity array with identical $a/d$ ($d/a = 1/2$). The solid lines are the Lorentz fitting results corresponding to the experimental results with hollow squares.

The position of transmission extinction is ~780 cm$^{-1}$ in fabricated devices with different diameter/periodicity ratios (Figure 6a), and a tunable resonant plasmonic peaks can be achieved readily by reconstructing the periodicity of nanocavity array, as indicated by equation (3). Figure 6c shows the extinction spectra for seven different nanocavity arrays with identical $d/a$ (equals 1/2) but different periodicities $a$ (100 to and 220 nm). As expected, the resonance frequencies can be tuned from ~1000 to ~720 cm$^{-1}$.

As a key feature of graphene, the ability to externally modulate its carrier concentration, such as electrostatic gating, allows a dynamical tunability of plasmonic resonance, which are impossible in noble metals. For our device, we further characterized the response in infrared of a same device ($d$ = 90 nm and $a$ = 180 nm) under seven different bias gate voltages (−4.0 to +0.8 V). The measured transmission extinction spectra are shown in Figure 7a and the scheme of applied voltage is shown in Figure 4a. There is a blue-shift in the resonant frequencies with increasing reverse gate voltages. The positions of resonant peaks can be easily tuned from ~560 to ~780 cm$^{-1}$ by varying the gate voltage from 0 to −4.0 V. When we applied a small forward gate voltage (+0.8 V), the resonant frequency of device has a red-shift due to the $p$-doped graphene of our device. Finally, we discuss the relationship between the resonant frequency $\omega_p$ and graphene carrier density $n$. Figure 7b shows the resonant frequency (extracted from Figure 7a) as a function of graphene carrier density. We observe that the resonant frequency changes non-linearly with $n^{1/4}$ and behaves a high dependence on the applied gate volate.

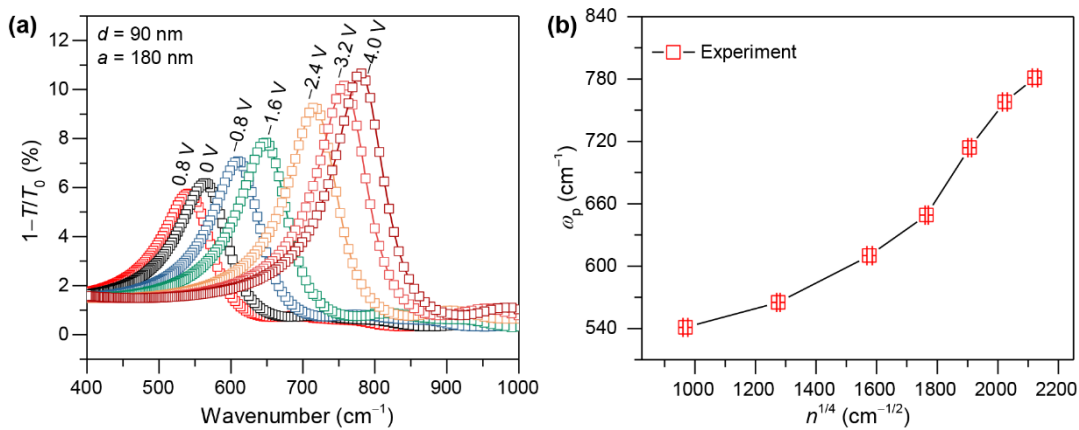

**Figure 7. Gate-tunable infrared filters.**
(**a**) Tuning of infrared extinction spectra by varying the gate voltage with identical geometry ($d$ = 90 nm, $a$ = 180 nm). (**b**) Correspongding plasmonic resonance frequency as a function of graphene carrier density modulated by gate voltage. Error bars indicate the S.D. of the calculated graphene carrier density within each unit.

## 3. Conclusion

We have observed the tunable infrared response in monolayer graphene integrated with patterned ferroelectric BFO thin film with nanocavity array. The carrier density of graphene can be spatially controlled with nondamaging technique and readily modulated by dynamically varying the applied gate voltage. The micromagnetic simulations indicate that plasmon waves were mainly localized at the edges of ferroelectric nanocavity, rather than propagating through the whole graphene sheet. The experimental results reveal that the resonant frequency can be easily tuned in infrared within a wide range from ~720 to 1 000 $cm^{-1}$ by reconstructing the ferroelectric nanocavity array and from ~540 to ~780 $cm^{-1}$ by changing the applied gate voltage. Our results demonstrate that the periodic ferroelectric nanocavity can be used to excite and control graphene plasmon waves with geometry and electrical current, and hence offer great versatility and the possibility to plasmonic devices.

## 4. Methods

### 4.1. Growth and patterning of BFO film

We first deposited a 25-nm-thick LSMO layer, as bottom conductive electrode, on STO substrate with (001)-orientation. After that, 25-nm-thick BFO layer was epitaxially grown. For the epitaxial growth of both BFO and LSMO thin films, a pulsed laser of 248 nm wavelength laser with KrF excimer was used. The repetition rate and energy density of the pulsed laser were 5 Hz and ~1.5 J $cm^{-2}$, respectively. An atmosphere with 0.2 mbar oxygen pressure and 700 °C thermal temperature were employed during both the BFO and LSMO films deposition. The BFO film was patterned to nanocavity arrays using a reactive ion etching method combined with a standard electron beam lithography technique before polarization. The etched nanocavity penetrated the BFO layer.

### 4.2. Device fabrication

The patterned ferroelectric cavity arrays were switched to uniformly downward polarization using a novel water-printing technique.[32] The whole BFO film was switched to downward polarization after exposing the film to the acidic aqueous solution (pH = 3). Deionized water was used to remove the acidic residue on the surface of ferroelectric films. Then, the CVD-grown monolayer graphene (SixCarbon Technology Shenzhen, China) layer was transferred onto the patterned BFO film with cleaned surface using an improved wet method reported previously.[37] After naturally drying, the prepared devices were placed into an oven for annealing 2 h at 200 °C under 100 sccm Ar:$H_2$ (volume ratio of 9:1) atmosphere. This strategy can improve the contact quality and remove the adhesive residue of devices.

### 4.3. Characterization and measurements

PFM characterizations were performed using an Infinity Asylum Research AFM (Oxford Instruments plc, UK) under ambient conditions. High resolution crystal characterizations of BFO films were performed using a transmission electron microscope (JEM 2100F, JEOL, Japan) operated at 200 keV. The optical images were obtained using a microscope (BX51M, OLYMPUS, Japan). High resolution device components were performed using a scanning electron microscope (JSM 7500F, JEOL, Japan) operated at 15 kV. Raman shifts and maps were detected using a tip enhanced Raman scattering (TERS) technique. For our experiments, the TERS system is fully equipped by integrating a scanning probe microscope with Raman micro-spectrometer (HORIBA, Japan). The wavelength of excitation laser in Raman characterizations is 532 nm. The laser power was set below 1 mW to avoid the laser-induced heating. The spectral transmission data were collected using a Fourier transform infrared microscopy system (Spotlight 200i, PerkinElmer, USA). All the characterization and measurements were performed at room temperature.

### 4.4. Numerical calculation

The dynamical conductivity of graphene was calculated using the random-phase approximation combined with the Kubo formula[40, 41] consisting of interband and intraband contributions as equation (4). The intraband part $\sigma_{intra}$ likes Drude mode, as shown in equations (5) and (6),

$$\sigma_g = \sigma_{intra} + \sigma_{inter,1} + i\sigma_{inter,2} \tag{4}$$

$$\sigma_{intra} = \sigma_0 \frac{4\mu_c}{\pi} \frac{1}{\hbar\tau_1 - i\hbar\omega} \tag{5}$$

$$\sigma_0 = \frac{\pi e^2}{2h} \tag{6}$$

where $\tau_1$ is the intraband relaxation rate and and $\mu_c$ is the chemical potential of graphene (namely the Fermi level. The interband part of graphene $\sigma_{inter,1}$ and $\sigma_{inter,2}$ can be obtained by following equations (7) and (8), respectively.

$$\sigma_{inter,1} = \sigma_0 \left( 1 + \frac{1}{\pi}\arctan\frac{\hbar\omega - 2\mu_c}{\hbar\tau_2} - \frac{1}{\pi}\arctan\frac{\hbar\omega + 2\mu_c}{\hbar\tau_2} \right) \tag{7}$$

$$\sigma_{inter,2} = -\sigma_0 \frac{1}{2\pi} \ln \frac{(2\mu_c + \hbar\omega)^2 + \hbar^2\tau_2^2}{(2\mu_c - \hbar\omega)^2 + \hbar^2\tau_2^2} \tag{8}$$

where $\tau_2$ is the interband relaxation rate.

For our designed plasmonic devices consisting of graphene and periodic BFO nanocavity, the near-field distribution were simulated using the finite element method. We set graphene a surface current with non-thickness. The ununiform conductivity of graphene was caculated following the method described as above mentioned. Periodical conditions were implemented on the double sides of our simulated region. The geometry of our simulated devices corresponds to the experimental results. The graphene chemical potentials were estimated using the shifts of Raman G-peak positions.

**Acknowledgements**

J.G. thanks Q. Mu for his great support of PFM characterization. The authors thank the CESHIGO Tech. for the high-resolution TEM characterization. J.G., L. Lin and S. Li thank Ms. X. Lai from the Sichuan University for the discussion and supports of the Raman measurements. J.G., L. Lin and S. Li thank Pro. J. Zhang, Dr. Y. Zhang and Dr. Y. Tian from the Beijing Normal University for the discussion and supports of the switching of ferroelectric domain and crystal structures of BFO thin films. This work was financially supported by the National Natural Science Foundation of China (No. 61971108).

**Conflict of Interest**

The authors declare no conflict of interest.

**Data Availability**

The data that support the findings of this study are available from the corresponding author upon reasonable request.